# The Order Is the Guarantee: Verifier-Budgeted Code Deletion with Static-First Learned Proposals

Ruitong Li[1,*] Binjie Guo[2,*] Aisheng Mo[2] Guowei Su[2]
Han Wang[3] Jie Li[4] Ru Zhang[2]

[1]University of Hong Kong [2]Zhejiang University
[3]Dalian University of Technology [4]Independent Researcher

[*]These authors contributed equally.

**Abstract**

Frontier coding models now match or exceed strong human reference points on programming benchmarks, yet benchmark success does not imply maintainable software. Prompt-driven "vibe coding" is additive: new branches, guards, and fallbacks accumulate faster than obsolete logic is removed. We study the inverse problem—how an AI system should remove code when execution-verification capacity is finite. We formulate redundant-code reduction as *proposal scheduling*: a ranker orders single-statement deletion candidates, an execution suite accepts the first candidate that passes, and a budget bounds how many candidates may be tested. Our central observation is that candidate *order*, not model confidence, is the control surface a deployment can reason about. DelScout instantiates two schedules. Given representative target-domain validation, a five-slot budget spends three slots on deterministic shortest-first candidates and two on complementary learned candidates; across nine MBPP replications with 0.5B, 0.6B, and 8B rankers this raises verified-deletion coverage by 9.5% relative (+6.7 accepted tasks) while consuming slightly fewer verifier calls than the matched static baseline. Without such validation the same rankers can lose coverage under shift, so we instead evaluate the complete static prefix first and append learned candidates only afterwards; for a deterministic verifier this makes coverage and character reduction non-decreasing by construction, at a measured 4.8–62.5% increase in verifier calls. MBPP+ then erases the in-domain advantage, showing that scheduling governs search while the test suite alone governs what "preserving behavior" means. The result is an auditable division of labor: models widen the search for removable code, order bounds the damage a mis-ranked proposal can do, and execution retains authority over every committed deletion.

# 1 Introduction

Large language models have transformed programming from token-level completion into end-to-end problem solving. Systems such as AlphaCode and DeepSeek-R1 now match or surpass strong human baselines on competitive-programming benchmarks, while contemporary code models tackle increasingly diverse generation tasks ranging from repository-level editing to full-feature implementation (Li et al., 2022; DeepSeek-AI, 2025; Hui et al., 2024; Jimenez et al., 2024). This progress, however, shifts the bottleneck. Producing yet another working implementation is becoming cheap; establishing that every generated branch, helper function, import statement, and compatibility layer remains genuinely necessary is not.

This imbalance is most evident in project-scale "vibe coding," where developers prompt, patch, and regenerate until tests pass. Each iteration tends to add a local fix while preserving earlier scaffolding, so abandoned alternatives persist and responsibilities become duplicated across edits. The program remains functional, but redundant code expands the review surface, obscures invariants, and solidifies into technical

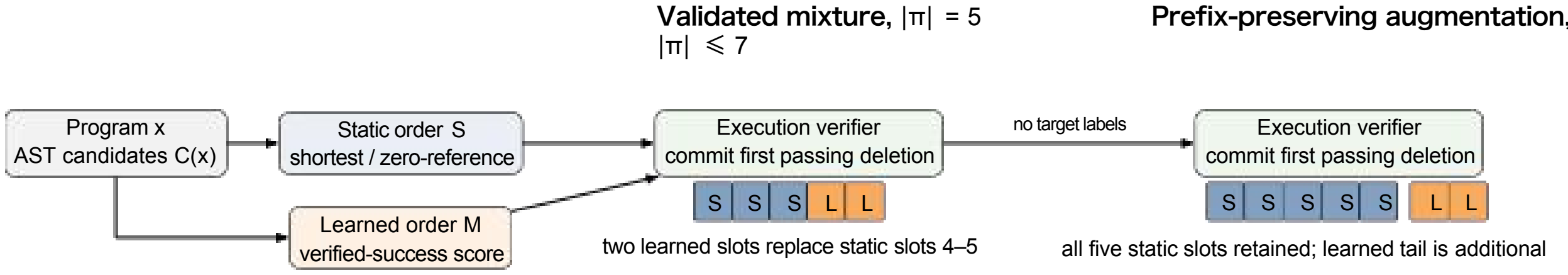

Figure 1: DelScout separates proposals from acceptance. The *validated mixture* spends a validated five-slot budget on a three-candidate static core plus two complementary learned candidates. *prefix-preserving augmentation* keeps the entire five-candidate static ranking and appends learned candidates only after static failure, so a learned ranking can never displace a static success. S and L denote static and learned proposals.

debt—a smell that developers recognize but rarely remove (Romano et al., 2020; Kruchten et al., 2012). Current generation and repair benchmarks reward code that passes tests; almost none reward a system for identifying what should no longer exist. We study this inverse problem as *verifier-backed code deletion*: propose a statement to remove, execute the relevant tests, and commit only if all deletions pass.

Deletion is intrinsically asymmetric. A useful proposal reduces maintenance burden, whereas a wrong deletion silently removes required behavior—which is why program reducers have always been organized around an executable oracle (Zeller and Hildebrandt, 2002; Regehr et al., 2012; Misherghi and Su, 2006; Sun et al., 2018). Sound static analysis captures the safe half of this problem: compilers and linters remove code they can prove unreachable or unused (Knoop et al., 1994). The hard half is code that is reachable and referenced yet behaviorally subsumed by the surviving implementation—a redundant guard, a duplicated normalization step, an obsolete fallback. Judging those requires contextual understanding, which is where a learned ranker can help—and where a model score is emphatically not a proof.

Once execution serves as the acceptance authority, the real design variable becomes the *schedule*: which proposals a finite verification budget is allowed to evaluate. This reframing matters because it clarifies what can and cannot be guaranteed. Learned candidates may complement a static ranking, but under a fixed budget they can also displace static candidates that would have passed under unknown distributional shift. Ordering, by contrast, is free and fully controllable. Our system, DelScout (Figure 1), lets the model order candidates while the execution suite retains veto power, and selects between two schedules based on the evidence actually available:

- When representative target-domain validation exists, the *validated mixture* allocates three of five verification slots to deterministic shortest-first candidates and the remaining two to the highest-ranked learned candidates absent from that core. Validation, not intuition, licenses this exchange.
- When such validation does not exist, *prefix-preserving augmentation* evaluates the *complete* five- candidate static ranking first and appends at most two learned candidates only if all five static candidates fail. Because a learned proposal never occupies a static slot, augmentation can add an accepted deletion but can never revoke one. The guarantee follows from ordering, not from calibration.

Our contributions are as follows. (i) We formulate redundant-code reduction as proposal scheduling under a finite execution-verification budget, cleanly separating learned search from the acceptance boundary. (ii) We introduce a target-validated static–learned mixture and a prefix-preserving augmentation policy whose non-decreasing coverage and character reduction follow from candidate order alone. (iii) We report nine replications across 0.5B–8B rankers, objective controls that rule out a privileged training loss, budget and verifier-cost accounting, cross-benchmark shift analysis, and stronger-verifier tests that locate precisely where the method stops transferring.

## 2 Related Work

**Program reduction and repair.** Reduction has always coupled candidate removal with an executable notion of preservation. Delta debugging and hierarchical delta debugging isolate failure-inducing fragments through repeated tests (Zeller and Hildebrandt, 2002; Misherghi and Su, 2006); C-Reduce and Perses add specialized passes and grammar guidance for efficiency and syntactic validity (Regehr et al., 2012; Sun et al., 2018). Generate-and-validate repair adopts the same split between search and acceptance: GenProg, Prophet, Angelix, TBar, SequenceR, and Recoder vary the proposal mechanism while tests judge patches (Weimer et al., 2009; Goues et al., 2012; Long and Rinard, 2016; Mechtaev et al., 2016; Liu et al., 2019; Chen et al., 2021b; Zhu et al., 2021). We inherit the verifier-centered principle but study redundancy removal and, unlike reducers that iterate to a local minimum, the allocation of a small fixed proposal budget across heterogeneous rankers.

**Dead code and technical debt.** Compiler dead-code elimination and modern linters delete what they can *prove* unnecessary, such as unreachable blocks and unused bindings (Knoop et al., 1994). Soundness is their strength and their ceiling: a reachable, referenced statement whose behavior is subsumed elsewhere is invisible to them. Empirical work reports that such residue is widespread, is perceived as a bad smell that impairs comprehension, and is only weakly correlated with where maintenance effort is actually spent (Romano et al., 2020; Eder et al., 2012; Kruchten et al., 2012; Fowler, 1999). Our measurements make the boundary concrete: on BigCodeBench-Hard the strongest static ranking succeeds mainly through unused imports (61 of 89 accepted deletions), whereas the learned tail is dominated by expressions and conditionals. Learning is therefore not an alternative to sound analysis; it extends the reachable candidate set past the point where proof is available, which is exactly why acceptance must stay with execution.

**Neural code models and execution feedback.** Code models have progressed from representation learning to competitive generation and repair (Feng et al., 2020; Guo et al., 2021; Wang et al., 2021, 2023; Fried et al., 2023; Li et al., 2023; Hui et al., 2024; Li et al., 2022; DeepSeek-AI, 2025), and adaptation studies show that modest fine-tuning can reprioritize useful edits (Silva et al., 2024; Hu et al., 2022; Xia and Zhang, 2022). Execution feedback has been used as reward, verification, test synthesis, and iterative repair (Le et al., 2022; Ni et al., 2023; Chen et al., 2023, 2024). HumanEval and MBPP established functional evaluation, while EvalPlus, DS-1000, BigCodeBench, SWE-bench, and automated test generation expose weak tests, library interactions, and repository-scale behavior (Chen et al., 2021a; Austin et al., 2021; Liu et al., 2023; Lai et al., 2023; Zhuo et al., 2025; Jimenez et al., 2024; Lukasczyk et al., 2022). These systems overwhelmingly optimize what to add or replace, and we treat each suite as a distinct operational specification rather than a pooled score. Our contribution is the scheduling view: execution gates every deletion, and prefix preservation bounds how a distribution-shifted learned ranking can disturb established static coverage.

## 3 Problem Formulation

Let $C(x)$ be the set of syntactically valid single-statement AST deletion candidates of program $x$, and let $x \setminus c$ denote $x$ with the line span of $c$ removed. A deterministic verifier $V$ for a domain returns 1 if $x \setminus c$ compiles and passes that domain's named execution suite, and 0 otherwise. A *proposal policy* $\pi$ returns an ordered duplicate-free list $\pi(x) = (c_1, \ldots, c_m)$ with $c_j \in C(x)$; the scheduler tests candidates left to right and commits the first that passes. Writing $\min\varnothing = \infty$, the accepted index is

$$\tau(\pi, x) = \min\{j \leq m : V(x \setminus c_j) = 1\}, \tag{1}$$

so $\tau = \infty$ means the policy commits no deletion and m is the policy's verifier-call bound for that program. Over an eligible corpus $\mathcal{D}$ (programs whose unmodified source passes the same suite), with $r(x, c)$ the fraction of source characters removed by c, we report

$$\mathrm{Cov}(\pi) = \tfrac{1}{|\mathcal{D}|}\left|\{x \in \mathcal{D} : \tau(\pi, x) < \infty\}\right|, \tag{2}$$

$$\mathrm{Red}(\pi) = \tfrac{1}{|\mathcal{D}|}\sum_{x:\,\tau(\pi,x)<\infty} r\big(x, c_{\tau(\pi,x)}\big). \tag{3}$$

Both sums range only over accepted tasks, so a task without a deletion contributes exactly zero and no undefined candidate is referenced. Coverage separates a ranker that finds many small deletions from one that finds fewer but larger removals, which Red captures.

**Schedules.** Let $S(x) = (s_1, s_2, \ldots)$ be a static order and $M(x) = (m_1, m_2, \ldots)$ a learned order over the same $C(x)$. Write $S_k(x)$ for the first k entries of $S(x)$, and let $M_B^{\setminus A}(x)$ be the first B entries of $M(x)$ that do not lie in the set A; $/\!\!/$ denotes ordered concatenation. The two schedules are

$$\pi_{\mathrm{mix}}(x) = S_3(x) \mathbin{/\!\!/} M_2^{\setminus S_3(x)}(x), \qquad |\pi_{\mathrm{mix}}| = 5, \tag{4}$$

$$\pi_{\mathrm{aug}}(x) = S_5(x) \mathbin{/\!\!/} M_B^{\setminus S_5(x)}(x), \qquad |\pi_{\mathrm{aug}}| \leqslant 5 + B. \tag{5}$$

De-duplication is defined against the static candidates the policy actually proposes—$S_3$ in (4) and $S_5$ in (5)—so each policy is a genuine list of distinct candidates and the two budgets are 5 and 5 + B with $B \leqslant 2$.

**Proposition 1 (prefix preservation).** For deterministic V and every $B \geqslant 0$, if $\tau(S_5, x) = j \leqslant 5$ then $\tau(\pi_{\mathrm{aug}}, x) = j$ and the committed candidate is the same $s_j$; if $\tau(S_5, x) = \infty$ then $\tau(\pi_{\mathrm{aug}}, x)$ is either $\infty$ or an index in the learned tail. Hence $\mathrm{Cov}(\pi_{\mathrm{aug}}) \geqslant \mathrm{Cov}(S_5)$ and, because $r \geqslant 0$, also $\mathrm{Red}(\pi_{\mathrm{aug}}) \geqslant \mathrm{Red}(S_5)$. The proof is immediate from (5): the first five entries of $\pi_{\mathrm{aug}}(x)$ are literally $S_5(x)$, and first-success stopping means a learned candidate is reached only when all five static candidates have already failed. The statement assumes a deterministic verifier and says nothing about behavior that V does not test.

**Proposition 2 (no fixed-budget dominance).** Fix a budget K and any schedule that omits some $s_j$ with $j \leqslant K$ from $S_K$. There is an instance on which $s_j$ is the unique candidate with $V(x \setminus s_j) = 1$: then $\tau(S_K, x) = j < \infty$ while the mixed schedule never proposes $s_j$ and returns $\tau = \infty$. No nontrivial fixed-K replacement therefore dominates the complete static prefix on every target distribution. Consequently a guarantee under arbitrary shift requires extra slots that preserve the prefix, whereas keeping the budget fixed requires empirical target-domain validation. The two schedules in (4)–(5) are exactly these two options, and Proposition 2 is why we never present the fixed-budget result as a universal claim.

# 4 Method

## 4.1 Why Two Proposal Families

AI-generated redundancy is not a single phenomenon. *Syntactic remnants* include imports, assignments, or helper functions left behind after the code path that required them has changed. *Control-flow redundancy* arises when successive prompts add guards or alternative branches already subsumed by the surviving implementation. *Compatibility scaffolding* comprises temporary fallback or migration logic that outlives the environment it originally protected. All three can appear locally plausible: the statement uses meaningful identifiers and resembles nearby code, even though removing it preserves all tested behavior.

These three forms reward different inductive biases, which is why we deliberately maintain two heterogeneous rankers rather than a single score. Shortest-first ranking suits syntactic remnants, since one-line assignments and imports carry low deletion risk. Zero-reference ranking uses identifier frequency to surface weakly connected statements. Neither can determine whether a conditional is behaviorally subsumed, because that judgment depends on surrounding control flow and task intent; a learned ranker supplies that contextual comparison. The rankers serve as complementary search operators over different manifestations of generated debt, not as interchangeable estimators of a single underlying quantity.

We restrict the action space to one complete AST statement. This unit covers the remnants above, maintains a one-to-one correspondence between a proposal and a source span, and makes every execution outcome attributable to a single edit. Multi-statement transformations could remove larger duplicated structures, but they entangle candidate generation with edit composition and multiply the possible explanations for a failure. Establishing the scheduling principle in the single-statement setting provides a controlled foundation for repository-scale extensions.

## 4.2 Candidate Space and Static Orders

We parse Python using the standard library AST and enumerate statement spans for imports, assignments, definitions, classes, control-flow statements, context managers, and expressions. A candidate is keyed by start line, end line, and node type; deleting it removes exactly the covered lines. Candidates that fail to compile are rejected and never counted as successes. The *shortest* order sorts by span length, then character length, then source position, then key. The *zero-reference* order first prefers statements whose identifiers occur least frequently in the file, then shorter spans. Both are deterministic and parameter-free, and which one is locally stronger is decided per domain without consulting learned test outcomes.

## 4.3 Learning Verified Utility, Not Surface Plausibility

The learned ranker is a sequence classifier over a prompt containing the candidate's node type and span, its text, and up to 24 numbered context lines on each side; the instruction asks whether deletion is safe because the statement is unused, and maps uncertainty explicitly to KEEP. Supervision comes only from verifier outcomes $y_{x,c} \in \{0, 1\}$ obtained by applying each candidate on the MBPP training range; test labels never enter prompts, training, or checkpoint selection. Let $z_\theta(x, c) \in \mathbb{R}$ be the classifier's success margin, that is the difference between its two output logits.

For each training task we form a group $G_x \subseteq C(x)$ containing every verified positive $P_x = \{c \in G_x : y_{x,c} = 1\}$ plus selected negatives up to $|G_x| \leq 8$, and we *discard tasks with* $P_x = \varnothing$, for which no ranking target exists. The listwise objective is then the negative log of the softmax mass that the model places on verified deletions,

$$\mathcal{L}_{\text{list}}(x) = -\log \frac{\sum_{c \in P_x} e^{z_\theta(x,c)}}{\sum_{c \in G_x} e^{z_\theta(x,c)}} \;\geq\; 0, \tag{6}$$

which is finite because $P_x \neq \varnothing$ and vanishes exactly when all mass sits on candidates the verifier accepts. Equation (6) expresses the deployment objective more directly than independent classification: a useful order needs one accepted candidate near the front, not calibrated probabilities for every AST node. Because (6) is invariant to the ranking among negatives, we add a small pointwise term that stabilizes the margin scale and retains a KEEP signal on unsuccessful candidates,

$$\mathcal{L}(x) = \mathcal{L}_{\text{list}}(x) + \frac{\lambda}{|G_x|} \sum_{c \in G_x} \text{BCE}\big(z_\theta(x, c), y_{x,c}\big), \tag{7}$$

with $\mathrm{BCE}(z, y) = -y \log \sigma(z) - (1 - y)\log(1 - \sigma(z))$, $\sigma$ the logistic function, and $\lambda = 0.1$ fixed for all runs. At deployment the softmax success score orders candidates, with the candidate key as a deterministic tie-break; no natural-language model output is ever executed.

Three controls change the support or weight of the same evidence while leaving (7) intact. *Exact-residual* training removes $S_3(x)$ from $G_x$, asking the model to rank only the opportunities that could occupy the learned tail. *Utility* training repeats each verified positive $3+\min(5, \lfloor |c|/80 \rfloor)$ times, where $|c|$ is its character length, so that larger useful deletions are favored by a factor between three and eight without letting one long span dominate optimization. *Task-marginal* training is stricter still: it keeps positive signal only for tasks with $\tau(S_3, x) = \infty$. That last objective matches marginal coverage conceptually but discards most positive evidence whenever static ranking is strong, so our experiments compare objectives instead of assuming that the most deployment-specific label is the easiest to learn.

This also clarifies what the classifier does not estimate. It does not predict semantic equivalence independently of a test suite, and it does not estimate the probability that an arbitrary project will tolerate a deletion. Its score is a task-conditioned ranking statistic learned from verifier outcomes in a named source domain, and shift can change both the candidate mix and the relation between surface context and tested behavior. We therefore use learned scores to order evidence-producing experiments, never as evidence themselves.

### 4.4 Choosing a Schedule

Complementarity, not dominance, is what makes a mixture worth its slots. For policies A and B, let $U_A = \{x : \tau(A, x) < \infty\}$. The relevant quantity is not $|U_B|$ but the decomposition $|U_B| - |U_A| = |U_B \setminus U_A| - |U_A \setminus U_B|$: new successes minus displaced ones. A ranker with lower standalone coverage remains valuable when its successes fall on the residual set $U_B \setminus U_A$, and a high standalone score does not justify replacement when $U_A \setminus U_B$ is large. This is why we report matched discordances rather than only aggregate counts, and why the *validated mixture* (Equation 4) keeps a static core instead of handing all five slots to the better standalone ranker.

The two schedules then represent two explicit contracts. The *validated mixture* holds resource use constant and accepts a measured risk of displacement in exchange for greater target-domain coverage; by Proposition 2, its empirical claim is confined to domains with representative validation data. *prefix-preserving augmentation* accepts a longer worst-case list—and, as we quantify below, more verifier calls—but makes prior coverage invariant to the learned order under a deterministic verifier. Neither contract makes a deletion semantically safe beyond the tests. Their value lies in making the source of uncertainty explicit: target evidence decides whether slots may be replaced, and the verifier decides whether any proposed edit may be committed.

### 4.5 Execution Procedure and Audit Trail

For each task, the scheduler materializes the ordered list before any patch runs. De-duplication is stable: when the learned and static orders name the same span, the earlier static occurrence owns the slot and the learned list advances. Every evaluation is transactional—copy the source, delete the candidate's complete line span, compile, execute the domain suite in an isolated process, and discard failures. The first passing patch is committed and later candidates are never tested. A program that fails its own baseline suite is ineligible and contributes no deletion.

## 5 Experimental Protocol

We ask four questions: whether one static-first architecture improves fixed-budget coverage across model scales and repeated runs; whether the gain comes from a special learning objective or from complementary scheduling; how learned replacement behaves under shift and whether prefix preservation repairs it; and which conclusions survive stronger public suites. MBPP (Austin et al., 2021) tasks 601–974 supply training labels, tasks 511–600 form the disjoint validation slice used for checkpoint selection, and tasks 1–500 are the held-out test, of which 499 or 500 are eligible depending on baseline executability. The same *validated mixture* is evaluated in nine independent runs spanning Qwen2.5-0.5B, Qwen3-0.6B, and Qwen3 -8B rankers with LoRA adapters. For shift, MBPP-trained 0.5B rankings are evaluated on DS-1000 (Lai et al., 2023), HumanEval (Chen et al., 2021a), and BigCodeBench-Hard (Zhuo et al., 2025) under their official execution oracles; *prefix-preserving augmentation* was designed after those results and is therefore exploratory there, after which it was frozen and evaluated once, without any adjustment, on HumanEval+ (138 eligible). MBPP+ (170 eligible) separately tests whether ordinary-MBPP deletions survive a stronger in-domain specification; both Plus suites are provided by EvalPlus (Liu et al., 2023). Per-run ledgers and supplementary analyses appear in the supplementary material.

| Backbone | Static | Mixture | Δ (range) |
|---|---|---|---|
| Qwen2.5-0.5B | 70.0 | 75.7 | +5.7 (+3, +7) |
| Qwen3-0.6B | 70.5 | 77.5 | +7.0 (+6, +9) |
| Qwen3-8B | 71.0 | 78.5 | +7.5 (+6, +9) |
| All nine runs | 70.4 | 77.1 | +6.7 |

Table 1: MBPP replications: mean accepted-task counts out of 499–500 eligible programs, at a matched five-proposal budget. Per-scale differences in parentheses are the minimum and maximum paired differences.

# 6 Results

## 6.1 Fixed-Budget Gains Replicate Across Model Scales

Table 1 aggregates the nine matched runs. Every paired comparison favors the mixture, with per-run gains of +3 to +9 accepted tasks, that is 4.3% to 12.7% relative. Mean coverage improves by 8.1%, 9.9%, and 10.6% relative for the 0.5B, 0.6B, and 8B rankers, an all-scale mean of 9.5% (70.4 → 77.1 accepted tasks, or 14.1% → 15.4% absolute coverage). The gain is positive in all nine matched runs and consistent across model scales.

## 6.2 Which Proposals Deserve the Slots

Figure 2 plots each individual ranking for K = 1 ... 5. The orders are far from interchangeable: candidate-feature logistic ranking is strongest at K = 1 (43 accepted versus 31 for shortest), listwise ranking is weakest there (27) yet overtakes shortest by K = 2, and utility ranking removes the most characters at every K $\geqslant$ 2. We keep this panel separate from the deployed comparison because the experiment recorded the preregistered K = 5 mixture rather than a retrospectively completed mixture curve.

At the matched budget of five proposals, Table 2 shows that all three mixtures improve coverage and character reduction over the five-candidate shortest-first baseline, and do so without spending more verification: the candidate-feature tail raises accepted deletions by 14.3% using 12 fewer verifier calls, and the utility-ranked tail reaches the strongest same-protocol point at 18.6% relative gain. Verified deletions per thousand verifier calls rise from 43.9 to 52.5, a 19.5% efficiency improvement, because a proposal that succeeds early also stops the schedule early. The utility point marks the attainable frontier; it does not establish utility weighting as universally preferable.

| Policy ($\|\pi\|$ = 5) | Accept | Calls | Acc./1k | Char. red. |
|---|---|---|---|---|
| Static shortest-5 | 70 | 1,593 | 43.9 | 2.11% |
| Mixture, candidate LR | 80 | 1,581 | 50.6 | 2.84% |
| Mixture, listwise | 78 | 1,586 | 49.2 | 2.99% |
| Mixture, utility | **83** | 1,581 | **52.5** | **3.06%** |

Table 2: Matched MBPP comparison at a common five-proposal budget over 500 tasks. Each mixture tests a three-candidate static core followed by two learned candidates. "Calls" is total verifier invocations and "Acc./1k" accepted deletions per thousand calls, so the mixtures improve coverage without buying it with extra verification.

Matched discordances make complementarity concrete. The utility-ranked mixture has 14 successes absent from the static baseline while the baseline has one absent from it. Against the candidate-feature mixture, the utility variant adds three tasks and loses none. The frontier is real; the identity of the best objective is not resolved by these data.

## 6.3 The Gain Is Architectural, Not an Objective Miracle

On 0.5B validation, global listwise and exact-residual training tie at a mean best score of 8.75 accepted tasks, and on the frozen test seeds available for both, global listwise averages 2.6% more accepted deletions than exact residual. The task-marginal objective peaked at 9 accepted validation tasks in all three of its seeds against a promotion gate of 10—the level the advanced arms reached—and was therefore never evaluated on test, a decision made before any test access. Its supervision is sparse by construction, since only tasks with a verified success outside the three-candidate static core contribute positive marginal signal. Together these controls rule out residual-only training as the source of the gain and leave the architectural explanation: static proposals cover easy candidates, learned proposals contribute a different order, and execution filters the incorrect ones.

Candidate composition shows the reallocation directly. At $K = 5$ the shortest order accepts 35 assignments but only 19 conditionals, with mean deletion length 1.39 lines. Candidate-feature ranking accepts 33 conditionals and 30 assignments at 1.62 lines, and listwise ranking accepts 32 conditionals while removing 3.16% of corpus characters. The candidate-feature mixture keeps all 35 assignment successes visible to its static prefix while raising conditional successes to 31 and mean deletion length to 1.58. The model is not reproducing "delete the shortest line" with noise; it spends its two slots on contextual control-flow candidates, which is precisely the region sound static analysis cannot certify—and also the region with more ways to be wrong, which is why execution remains mandatory. Character reduction confirms that the gain is not concentrated in trivia: every mixture improves on both axes of Table 2 simultaneously.

## 6.4 Under Shift, Preserve the Prefix

Model-only transfer is unreliable (Table 3). On DS-1000 learned coverage spans 9.35–9.70% against a 9.58% zero-reference baseline; on HumanEval it swings from 59.15% to 96.95% around a 93.29% baseline; on BigCodeBench-Hard it reaches only 41.22–50.00% against 60.14% for shortest-first. The learned orders do remove more characters when they succeed—2.61–3.07% versus 1.24% on BigCodeBench-Hard—so the failure is a coverage–reduction trade-off rather than uniform incompetence. Either way, a ranker that helps in-domain cannot justify replacing a complete static prefix under unknown shift, exactly as Proposition 2 predicts.

Prefix-preserving augmentation changes the operational conclusion without claiming the scores became calibrated. Learned replacement can push coverage below the static reference, whereas augmentation was non-degrading in all nine frozen replays and strictly positive in eight, improving mean coverage by 0.23, 2.24, and 3.38 percentage points on the three benchmarks. This is not free. Because the extra slots are only reached after five static failures, they cost 62.5%, 4.8%, and 24.8% more verifier calls respectively—the price of the guarantee, and on DS-1000 an unattractive one for a 0.23-point gain. The cost is explicit and quantifiable, and should be weighed against the coverage improvement in any deployment decision.

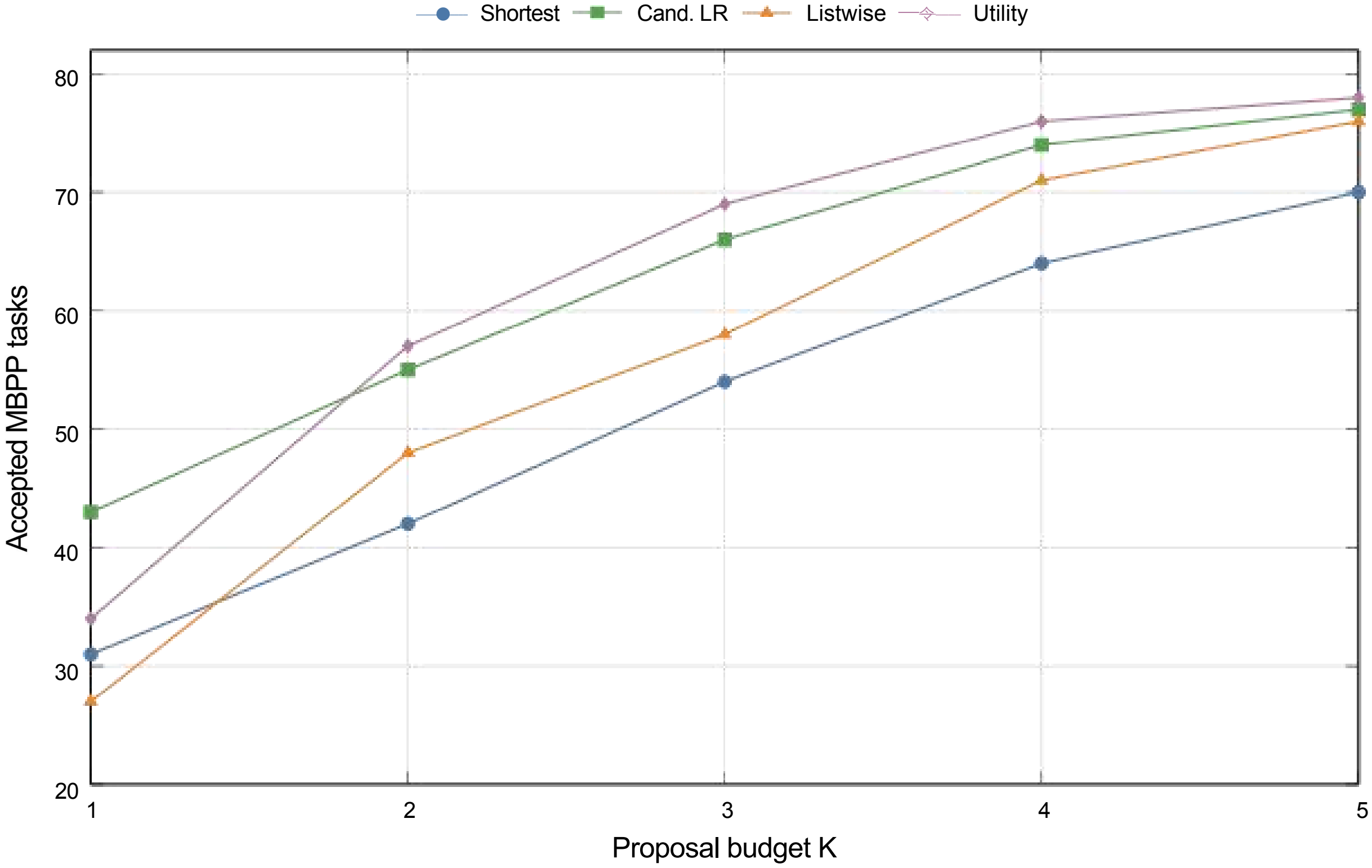

Figure 2: MBPP budget curves for the individual orders, not the mixtures. The rankings disagree most at small K, which is what a five-slot schedule has to arbitrate.

## 6.5 Stronger Tests Define the Safety Boundary

MBPP+ gives a deliberately adverse result. Among the 170 programs that pass the stronger suite before deletion, the five-candidate shortest-first baseline and all three static-first mixtures accept exactly 11 deletions, that is 6.47% coverage, and the mixtures raise base-only rejections—patches accepted by the original MBPP tests but rejected by MBPP+—from 3 to 7. The ordinary-MBPP advantage does not survive the stronger specification, and the mixtures' additional proposals fail disproportionately under it. "Verified" always means verified by the stated oracle.

HumanEval+ evaluates the frozen *prefix-preserving augmentation* without target-specific adjustment. The complete five-candidate static ranking accepts deletions for 92.75% of the 138 eligible programs; across three runs augmented coverage is 97.83%, 92.75%, and 94.93%, so two runs improve and one ties, at 4.3–5.7% extra verifier calls (Table 4). The replay reuses frozen standard-verifier ledgers, which do not expose every unsuccessful secondary candidate, so the augmented figures are conservative lower bounds rather than a complete two-candidate replay. These results test the ordering property of Proposition 1 under a stronger oracle and reinforce that "verified" always means verified by the stated specification.

| Dataset | Elig. | Static | Learned only | Augm. | Calls |
|---|---|---|---|---|---|
| DS-1000 | 866 | 9.58 | 9.35–9.70 | 9.82 | +62.5% |
| HumanEval | 164 | 93.29 | 59.15–96.95 | 95.53 | +4.8% |
| BCB-Hard | 148 | 60.14 | 41.22–50.00 | 63.51 | +24.8% |

Table 3: Accepted-deletion coverage (%) under distribution shift for the locally strongest static order (zero-reference on DS-1000 and HumanEval, shortest-first on BigCodeBench-Hard), the three learned-only rankings, and *prefix-preserving augmentation* (mean over the same three runs). The last column is the augmented policy's verifier-call overhead relative to the static order. This shifted-benchmark analysis is exploratory.

| Verifier | Policy | Coverage | Char. red. |
|---|---|---|---|
| MBPP+ | static shortest-5 | 6.47% | 0.76% |
| MBPP+ | mixtures (all three) | 6.47% | 0.78% |
| HE+ | complete static-5 | 92.75% | 56.28% |
| HE+ | augmented, 3 runs | 92.75–97.83% | 56.28–61.21% |

Table 4: Strong-verifier boundary on 170 MBPP+ and 138 HumanEval+ (HE+) eligible programs. The MBPP+ tie holds for all three mixtures, which also raise base-only rejections from 3 to 7.

# 7 Discussion

AI coding has fundamentally altered the economics of software creation. When frontier models produce benchmark-correct solutions at or above strong human reference points, generation ceases to be the scarce operation; understanding, consolidation, and removal become comparatively more valuable. Vibe coding sharpens this imbalance because each prompt adds a local solution while preserving earlier scaffolding. The resulting redundancy is not merely aesthetic: it expands review surface, multiplies the states future agents must reason about, and converts short-term generation speed into long-term debt. Deletion deserves to be a first-class AI coding capability rather than an occasional manual cleanup.

The design choice we defend most strongly is separating discovery from authority, and then being explicit about which evidence licenses which schedule. Learned rankers recognize contextual redundancy that static orders miss, but their scores are proposals, not correctness evidence. With representative validation data and a fixed five-slot budget, the *validated mixture* trades two static slots for contextual search; without such data, *prefix-preserving augmentation* tests the full static ranking first and consults learned candidates only after all five fail, so the learned tail can discover extra removals but can never suppress a deletion the established policy would have found. What decides between them is available target evidence and the required guarantee—not a vague judgment about how much to trust the model.

The abstraction extends past statement deletion. Repository agents and IDEs could use it to retire duplicated helpers after generation, simplify patches before review, remove compatibility scaffolding after migrations, or prioritize refactoring candidates in continuous integration: a conservative analyzer occupies a protected prefix, a learned model explores complementary candidates, and the project's own tests gate commitment. Larger models may improve the learned tail, but they cannot substitute for test adequacy, and our MBPP+ result offers the clearest demonstration of that limit. The practical opportunity is a maintenance loop in which AI systems not only generate software but continuously justify and remove what the project no longer needs.

# 8 Limitation

Our labels and targets are benchmark programs rather than maintenance histories, and the effect sizes are modest in absolute terms: coverage moves from about 14% to 15% of MBPP programs. Flaky or environment-dependent tests break the determinism that Proposition 1 assumes, so a deployment should pin dependencies, isolate execution, and repeat nondeterministic tests. Finally, deleting code can remove logging, monitoring, security checks, or rare behavior that no test exercises. Human review and adequate specifications remain necessary; DelScout must not be presented as a semantic proof, nor used to weaken safety-critical code merely because a benchmark-style verifier passed.

# 9 Conclusion

The next frontier of AI coding is not only generating more code but keeping generated software comprehensible as it evolves. We formulated that maintenance problem as verifier-backed deletion under a finite proposal budget and introduced DelScout, which combines deterministic static candidates with context-sensitive learned candidates while reserving acceptance for execution tests. A shared static-first mixture improves MBPP coverage across three model scales and nine runs without extra verification cost, and objective controls attribute the benefit to complementary scheduling rather than a privileged training loss. Under shift, replacing static candidates can fail, and retaining the complete static prefix restores non-decreasing coverage by construction—at a verifier-call cost we measure rather than hide. Stronger Plus suites then mark the boundary: scheduling governs search, but only the specification governs what "preserving behavior" means.

The governing principle is deliberately simple. Let models broaden the search for unnecessary code, let order constrain their failure modes, and let project evidence decide what may be deleted. Reliable AI coding is not complete when a system can write a working program; it is complete only when the system can also help that program stay no larger than it needs to be.

# Appendix

## A Experimental Configuration

MBPP (Austin et al., 2021) tasks 601–974 provide training labels, tasks 511–600 form the disjoint validation slice used for checkpoint selection, and tasks 1–500 form the held-out test; 499 or 500 programs are eligible depending on baseline executability. The same static-first mixture—three deterministic shortest-first candidates followed by two learned candidates absent from that core, within a five-proposal budget—is evaluated in nine independent runs across Qwen2.5-0.5B, Qwen3-0.6B, and Qwen3-8B rankers. Distribution-shift evaluation uses DS-1000 (Lai et al., 2023), HumanEval (Chen et al., 2021a), and BigCodeBench-Hard (Zhuo et al., 2025) with their official execution suites; MBPP+ and HumanEval+ from EvalPlus (Liu et al., 2023) provide stronger public specifications.

Each ranker is a sequence classifier with two output heads and a LoRA adapter of rank 8 and scaling 16, dropout 0.05, applied to the query and value projections, with the classification head also trainable. Training uses AdamW at learning rate $2 \times 10^{-5}$, gradient clipping at norm 1.0, gradient accumulation over four task groups, a maximum prompt length of 768 tokens, and four epochs, of which validation selects one. Task groups contain every verified positive plus negatives up to a group size of eight; negatives are ordered to prefer the node types that appear among that task's positives, then shorter spans, so the listwise contrast is not dominated by trivially dissimilar candidates. The prompt contains the candidate's node type and span, its text, and up to 24 numbered context lines before and after, and the instruction explicitly maps uncertainty to KEEP. At deployment the softmax success score orders candidates with the candidate key as a deterministic tie-break; no natural-language model output is executed.

Selected 8B checkpoints were rescored twice over all 4,863 candidates and the resulting candidate orders were bitwise identical (SHA-256 recorded per seed). Validation selections, task-level outcomes, and analysis inputs are retained in machine-readable form, so every reported aggregate can be regenerated without exposing environment-specific paths or identifiers.

## B Formal Interpretation of the Scheduler

Let $\mathcal{D}$ be the eligible program set. For each $x \in \mathcal{D}$ a policy returns an ordered, duplicate-free list $\pi(x) = (c_1, \dots, c_m)$; candidate $c_j$ deletes one complete statement, producing $x \setminus c_j$, and the deterministic verifier $V$ returns 1 exactly when that patched program compiles and passes the named suite. With $\min\varnothing = \infty$, define the passing-index set and the accepted index

$$J_V(\pi, x) = \{j \in \{1, \dots, m\} : V(x \setminus c_j) = 1\}, \tag{8}$$

$$\tau_V(\pi, x) = \min J_V(\pi, x), \tag{9}$$

so $\tau_V(\pi, x) = \infty$ means the policy commits no deletion. Verified-deletion coverage is

$$\mathrm{Cov}_V(\pi; \mathcal{D}) = \frac{1}{|\mathcal{D}|} \sum_{x \in \mathcal{D}} \mathbf{1}[\tau_V(\pi, x) < \infty]. \tag{10}$$

Let $\ell(x)$ be the number of source characters and $r(x, c) = (\ell(x)-\ell(x\setminus c))/\ell(x)$. Per-task character reduction is defined without referring to an undefined $c_\infty$:

$$R_V(\pi, x) = \begin{cases} r(x, c_{\tau_V(\pi,x)}), & \tau_V(\pi, x) < \infty, \\ 0, & \tau_V(\pi, x) = \infty, \end{cases} \tag{11}$$

and the dataset-level reduction is $\mathrm{Red}_V(\pi; \mathcal{D}) = \frac{1}{|\mathcal{D}|} \sum_x R_V(\pi, x)$. A failed task therefore contributes zero and a successful task contributes only its first verifier-passing deletion.

Write $S_k(x) = (s_1, \ldots, s_k)$ for the first k entries of the static order and $M_B^{\setminus A}(x)$ for the first B entries of the learned order that do not belong to the candidate set A; $/\!/$ denotes ordered concatenation. The target-validated fixed-budget policy and the shift policy are

$$\pi_{mix}(x) = S_3(x) /\!/ M_2^{\setminus S_3(x)}(x), \qquad |\pi_{mix}| = 5, \tag{12}$$

$$\pi_{aug}(x) = S_5(x) /\!/ M_B^{\setminus S_5(x)}(x), \qquad |\pi_{aug}| \leqslant 5 + B. \tag{13}$$

De-duplication is taken against the static candidates each policy actually proposes: $S_3$ for the mixture, whose static slots 4 and 5 are the ones being exchanged, and $S_5$ for augmentation. The first policy replaces two static slots only when target-domain validation supports the exchange; the second never replaces a static proposal and only adds search beyond the complete prefix.

**Proof of prefix preservation.** The first five entries of $\pi_{aug}(x)$ are exactly $S_5(x)$. If $\tau_V(S_5, x) = j \leqslant 5$, then evaluation of $\pi_{aug}$ tests the identical candidates $s_1, \ldots, s_j$ in the identical order, all of $s_1, \ldots, s_{j-1}$ fail as before because V is deterministic, and $s_j$ is committed; hence $\tau_V(\pi_{aug}, x) = j$ with the same accepted candidate. If $\tau_V(S_5, x) = \infty$, augmentation either also fails or succeeds somewhere in its learned tail. In both cases

$$\mathbb{1}[\tau_V(\pi_{aug}, x) < \infty] \geqslant \mathbb{1}[\tau_V(S_5, x) < \infty], \tag{14}$$

and averaging the pointwise inequality gives $\mathrm{Cov}_V(\pi_{aug}) \geqslant \mathrm{Cov}_V(S_5)$. Because static successes retain the same accepted candidate and every new learned success has $r(x, c) \geqslant 0$, the same argument gives $\mathrm{Red}_V(\pi_{aug}) \geqslant \mathrm{Red}_V(S_5)$. Both statements require deterministic execution and constrain only behavior represented by V. They say nothing about the number of verifier calls, which strictly increases on tasks where the static prefix fails; that cost is reported in the main paper.

**Why replacement needs validation.** Suppose a fixed-budget mixture omits some $s_j \in S_K$. Consider an instance on which $V(x \setminus s_j) = 1$ while every retained static candidate and every learned candidate fails. Then $S_K$ succeeds and the mixture does not, so no nontrivial replacement policy dominates the complete static prefix on every target distribution. What governs the observed gain is instead complementarity: with $U_\pi = \{x : \tau_V(\pi, x) < \infty\}$,

$$|U_\pi| - |U_S| = |U_\pi \setminus U_S| - |U_S \setminus U_\pi|. \tag{15}$$

The first term counts new successes and the second counts displaced static successes. Prefix-preserving augmentation forces the second term to zero by construction, whereas a fixed-budget mixture must estimate both terms on representative validation data.

# C Complete MBPP Replications

The per-run detail matters for interpretation. The directional result is consistent across all nine runs. Shortest-first itself accepts 70 or 71 tasks depending on which frozen run is read, because one program's eligibility depends on baseline executability; all comparisons are therefore paired within a run.

# D Distribution Shift, Per Seed

Two properties of this replay bound its interpretation. First, the augmented arm is reconstructed from frozen official-verifier ledgers: a learned candidate is credited only if the learned-only run had already accepted it within its own first B attempts, so accepted counts are lower bounds. Second, the replay charges up to B learned attempts without removing overlaps with the already-failed static prefix. Such an overlap cannot create an accepted deletion—the same candidate under the same deterministic verifier fails again—so the guarantee is unaffected, and the reported call overheads are upper bounds on what a de-duplicated implementation would spend.

| Backbone | Objective | Static | Mixture | $\Delta$ |
|---|---|---|---|---|
| Qwen2.5-0.5B | residual | 70 | 77 | +7 |
| Qwen2.5-0.5B | residual | 70 | 73 | +3 |
| Qwen2.5-0.5B | residual | 70 | 77 | +7 |
| Qwen3-0.6B | listwise | 71 | 77 | +6 |
| Qwen3-0.6B | listwise | 71 | 80 | +9 |
| Qwen3-0.6B | listwise | 70 | 76 | +6 |
| Qwen3-0.6B | listwise | 70 | 77 | +7 |
| Qwen3-8B | residual | 71 | 80 | +9 |
| Qwen3-8B | residual | 71 | 77 | +6 |
| All nine | — | 70.4 | 77.1 | +6.7 |

Table 5: Every unlocked replication, in accepted-task counts on the held-out MBPP test. All nine paired differences are positive.

| Dataset | Static | Learned only | Augmented |
|---|---|---|---|
| DS-1000 (866) | 83 | 81 / 82 / 84 | 85 / 84 / 86 |
| HumanEval (164) | 153 | 159 / 97 / 148 | 161 / 153 / 156 |
| BCB-Hard (148) | 89 | 61 / 68 / 74 | 93 / 95 / 94 |

Table 6: Accepted deletions per seed under shift, with eligible programs in parentheses. Static is the locally strongest static order (zero-reference on DS-1000 and HumanEval, shortest-first on BigCodeBench-Hard). Verifier calls rise from 1,951 to 3,171 on DS-1000, from 413 to 433 on HumanEval, and from 468 to 584 on BigCodeBench-Hard, i.e. +62.5%, +4.8%, and +24.8% on average.

The guarantee also assumes deterministic execution. Under a flaky verifier the same static candidate can change outcome between schedules even when its position is preserved, so a practical implementation should pin dependencies, isolate execution, and repeat nondeterministic tests.

# E Stronger Verifiers

On the MBPP+ intersection (170 programs passing the stronger suite before deletion) shortest-first and all three static-first mixtures accept exactly 11 deletions. The mixtures record seven base-only rejections against three for shortest-first, and the learned-only orders accept 10 or 11 with six base-only rejections. The ordinary-MBPP advantage therefore does not survive the stronger specification.

HumanEval+ contributes the 138 of 164 programs that pass both the base and Plus suites before deletion. The complete five-candidate zero-reference ranking is evaluated first, and at most two learned proposals are considered only after all five fail. Static coverage is 128/138 (92.75%); the three augmented runs obtain 135, 128, and 131 accepted deletions (97.83%, 92.75%, 94.93%) with character reduction 61.21%, 56.28%, and 58.88% against 56.28%, at 363, 368, and 367 verifier calls against 348.